\documentstyle[epsfig]{aipproc}

\begin{document}
\title{UBVRI photometry of Southern Sky\\ BL Lacs}

\author{R. R. Sefako$^1$, O. C. de Jager$^1$ and H. Winkler$^2$}
\address{$^1$Space Research Unit, Potchefstroom University, Potchefstroom,
2520, South Africa\\ 
$^{2}$ Dept of Physics, Soweto Campus, Vista University, Johannesburg, South
Africa}

\maketitle

\begin{abstract}
Seventeen southern sky BL Lacs were observed in UBVRI
using the CCD Camera on the 1.0m
telescope at the South African Astronomical Observatory (SAAO) in Aug and 
Nov 1999. The analyses of all the seventeen sources are now complete,
and are available via anonymous ftp (ftp pukrs1.puk.ac.za/pub/Blazars). A 
few examples of our results are however 
given in this paper. Whereas PKS 2005-489 and 2155-304
appear to have been in a high state, PKS 0048-097 and PKS 0521-365 showed 
evidence of variability on a time-scale of a few days, with the amplitude
of variability increasing towards short wavelengths. This is consistent
with observations of gamma-ray BL Lacs, which show similar behaviour
in optical and X-rays.   
\end{abstract}

\section{Introduction}
BL Lacs are the only members of the population of extragalactic sources
which have been observed at TeV energies. They
emit their energy over a broad range of the spectrum (radio to 
$\gamma$-rays), with signatures of variability in optical, X-rays and 
$\gamma$-rays, which are large
compared to other AGNs. BL Lacs are highly polarized non-thermal sources
with compact flat-spectrum radio cores. Most BL Lacs were identified in 
radio and X-ray surveys. Whereas those identified in X-rays (XBLs) 
are mostly nearby, the radio selected BL Lacs (RBLs) have a larger distance
scale. Whereas a few XBLs were detected as TeV sources,
we find that the RBLs were mostly detected as EGRET sources
(see e.g. Lin et al. 1997, 1999).

The number of BL Lacs known presently are mostly northern sources, which 
constitute $\sim$75\% of all known BL Lacs. 
Convincing detections of VHE photons from 
BL Lacs have mainly been made from Northern Hemisphere sources, which include
Mrk 421 (Punch et al. 1992), Mrk 501 (Quinn et al. 1996) and 1ES 2344+514
(Catanese et al. 1998). TeV observations of Southern Sky sources mostly gave 
upper limits (e.g. Roberts et al. 1999). Besides PKS 2155-304, which has 
also been recently confirmed as a TeV source (Chadwick et al. 1999), 
none of the southern BL Lacs have been well
monitored. Therefore, it is justifiable to claim that there are a number
of BL Lacs that have not been identified from the southern skies, and probably 
with advances in VHE ground-based telescopes, some are likely to be identified
as TeV sources in future. Our objective is to monitor Southern Sky BL Lacs 
in optical, and such information can be combined with contemporary X-ray
observations for modeling purposes to identify further TeV candidates.

\section{Observations}
CCD observations in the UBVRI of several Southern BL Lacs were carried out 
with the SAAO 1.0m telescope in Sutherland during Aug and Nov 1999. A
number of E-regions standard stars were also observed for comparison and 
determination of apparent magnitudes of the target sources. The data were 
cleaned and flat-fielded using the IRAF image-processing software at SAAO in
Cape Town. Table~1 gives a list of BL Lacs that were observed. 

The data were analysed and the differential photometry of all seventeen 
sources have 
been obtained. We used the E-region stars and, in some cases, apparently  
non-variable stars which appear on the frame of the target source as standard
stars. If the 
stars on the frame do not have known published magnitudes, we used the 
E-regions
stars to determine their magnitudes which were eventually used to determine 
the differential photometry (as shown in Table~2). 
Atmospheric 
extinction and Galactic reddening corrections were not taken into account 
in our analysis, since the `standard' stars were mostly on the same frame as 
the target. The results of the rest of the observations will be released 
in the near future.  

\begin{table}
\begin{center}
\caption{A list of BL Lacs that were observed in Aug and Nov 1999.}
\begin{tabular}{l c l l c}
 & & & \\ 
Object&z& Time when observed& Filters& Frames\\ 
\hline
PKS 2005-489&0.071& 24 Aug - 5 Sep& UBVRI& 42\\
MH 2136-428&& 26 Aug - 5 Sep &BVRI& 20 \\
PKS 2155-304&0.117& 24 Aug - 5 Sep &UBVRI& 40\\
PKS 2254-204&& 24 Aug - 27 Aug &BVRI& 10\\
MS 2306-223&0.137& 25 Aug, 24 - 29 Nov&BVRI& 21 \\
PKS 2316-423&0.055& 25 Aug - 5 Sep, 26 - 29 Nov &BVRI& 36 \\
PKS 0048-097&$>$ 0.2& 25 Aug - 5 Sep, 28, 29 Nov &BVRI& 32 \\
PKS 0215+015&1.715& 25 Aug - 5 Sep, 27, 28 Nov &BVRI& 28 \\
PKS 0301-243&& 24 Aug - 5 Sep, 25 - 29 Nov &BVRI& 36 \\
PKS 0338-214&& 24 Aug - 5 Sep, 25 - 29 Nov &BVRI& 40 \\
EXO 0423-084&0.039& 24 Aug - 5 Sep, 27 - 29 Nov &BVRI& 36 \\
PKS 0521-365&0.055& 27 Aug - 5 Sep, 24 - 29 Nov &BVRI& 40 \\
PKS 0537-441&0.896& 01 Sep - 4 Sep, 24 - 29 Nov &BVRI& 28 \\
PKS 0548-322&0.069& 30 Aug - 2 Sep, 27 Nov &BVRI& 16 \\
PKS 0219-164&0.698& 25 Nov - 29 Nov &BVRI&12 \\
RXJ 0316-260&& 25 Nov - 29 N0v &BVRI& 12 \\
EXO 0556-383&& 27 Nov &BVRI&4 \\
\hline
\end{tabular} 
\end{center} 
\end{table}

\section{Results}
The UBVRI magnitudes of six BL Lacs are listed in Table 2 
 together with the catalogued V values of Padovani \& Giommi (1995).
Variability is evident for some of these sources. PKS 2005-489 seemed to be in
a high state, with its V-band flux about four times larger 
during Aug/Sep 1999 (in comparison to its
catalogued value), whereas the V-band flux from
PKS 2155-304 increased by 80\% during Sep 1999. 
Even though PKS 0048-097 and PKS 0521-365 
were close to their respective low states, they show evidence of variability 
on a time-scale of a few days (see Figure 1).
We noted however that $\Delta B > \Delta I$ for 
both PKS 0048-097 and 
PKS 0521-365, which means the amplitude of variability increases towards
higher frequencies.

\begin{table}
\begin{center}
\caption {Sample results of our data for the 2nd of Sep. 1999. V$^{p}$  
represents visual magnitudes published in Padovani \& Giommi (1995). 
Errors are estimated to be $\sim$ 0.05 mag.}
\begin{tabular}{l c c l c c c c}\\
 Object & MJD &      U&       B &      V& V$^{p}$&      R&     I\\
\hline 
PKS 2005-489       
&51423.978 & 12.55 &  13.11 &  12.80 &14.4 &12.41 & 12.00  \\
PKS 2155-304
&51423.993 & 12.52 &  13.12 & 12.84 &13.5 &12.54 & 12.13\\
PKS 2316-423
&51424.019 && 16.57 & 15.46 &14.5 &14.84 & 14.14\\ 
PKS 0048-097
&51424.040  && 16.9 &  16.45 &16.3 &16.06 & 15.50\\
EXO 0423-084
&51424.086 && 17.29 & 15.98 &15.9 &15.23 & 14.42\\
PKS 0521-365    
&51424.103 &&  15.61 & 14.92 &14.6 &14.38 & 13.74\\
\hline
\end{tabular}
\end{center}
\end{table} 

\begin{figure}
\centerline{
\epsfxsize=7.5cm
\epsffile{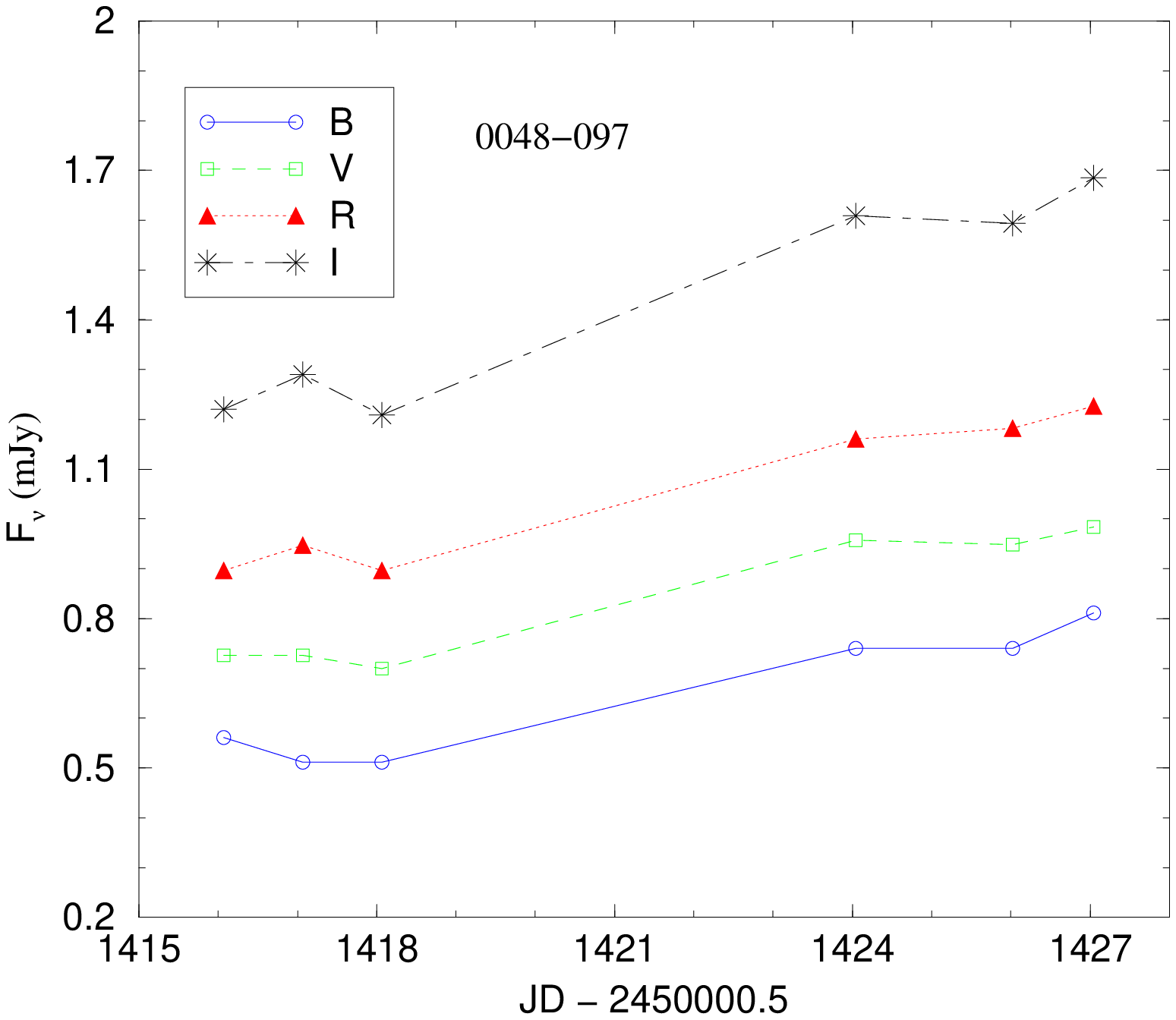}
\epsfxsize=7.32cm
\hspace*{-0.15cm}\epsfbox{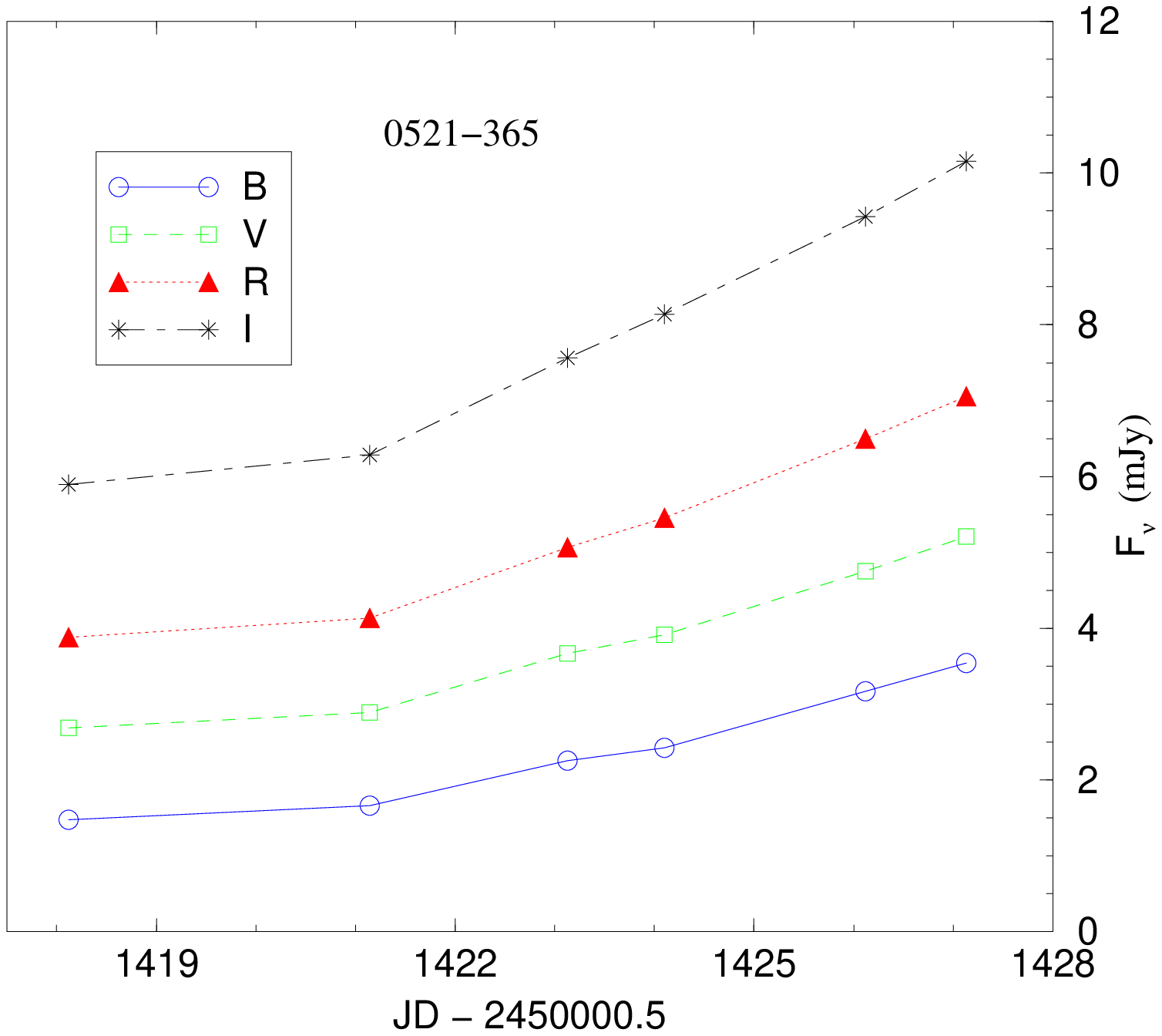} 
}
\caption{BVRI light-curves of PKS 0048-097 and PKS 0521-365. Vertical 
axes' labels are for PKS 0048-097 
(left) and PKS 0521-365 (right).Errors are less than $\sim$0.06 mJy.}
\end{figure}

In Fig. 2, we show the spectra 
of PKS 2005-489 and PKS 2155-304, with our UBVRI points included. The 
other data points (Infrared) for both sources are from the Catalog 
of Infrared Observations (NASA RF--1294, 3rd Edition). 

\begin{figure}
\centerline{
\epsfxsize=12.4cm
\epsffile{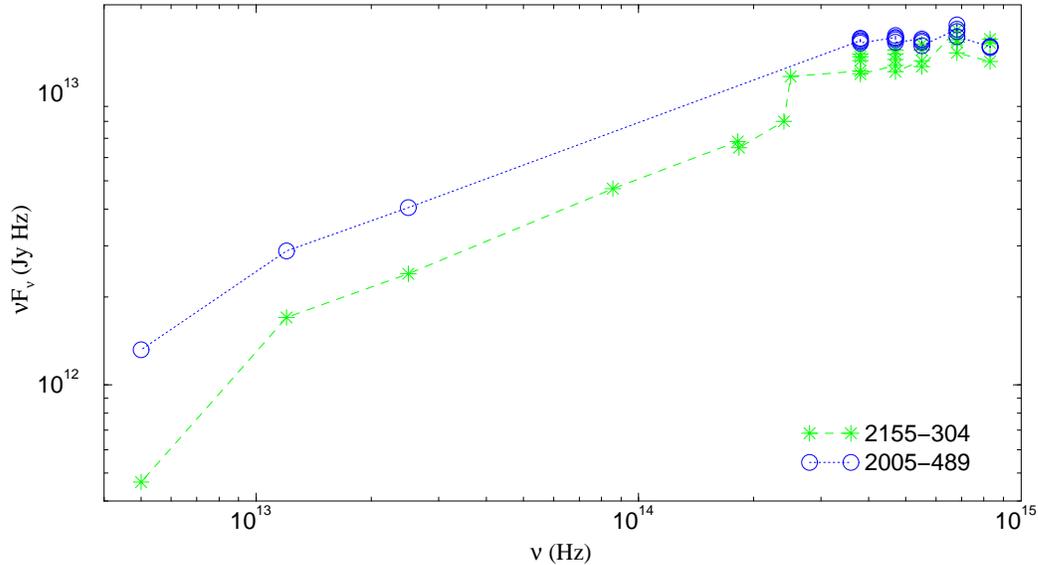}
}
\caption{SED of PKS 2005-489 and PKS 2155-304 from far the IR (60 $\micron$) 
to the visual band (0.36$\micron$). Optical data are from our analysed data.}

\end{figure}

\section{Conclusions}
Whereas BL Lac variability is a common feature,
our visual multicolor observations provide important information concerning 
the relative flux change vs. frequency. The observation of an increasing
amplitude of flux variability towards higher frequencies (in this case
for PKS 0048-097 and PKS 0521-365) should be combined with simultaneous
X-ray observations. This synchrotron feature in the optical/X-ray region 
of the spectrum should also imply a similar behaviour in the inverse Compton 
($\gamma$-ray)  component of the spectrum.

\section*{References}
\def\rf{\noindent\hangafter=1\hangindent=1truecm}

\rf 
Catanese, M., et al., 1998, ApJ, 501, 616

\rf
Chadwick, P. M., et al., 1999, ApJ, 513, 161  

\rf
Lin, Y. C., et al., 1999, ApJ, 525, 191

\rf
Lin, Y. C., et al., 1997, ApJ, 476, L11
 
\rf
Padovani, P., Giommi, P., 1995, MNRAS, 277, 1477

\rf 
Punch, M., et al., 1992, Nature, 160, 477

\rf 
Quinn, J., et al., 1996, ApJ, 456, L83

\rf
Roberts, M., et al., 1999, ASP Conference Series on {\it BL Lac Phenomenon}, 
Turku, Finland, ed. L. O. Takalo \& A. Sillanp\"{a}\"{a}, Vol. 159, 233

\end{document}